\documentclass[twocolumn,showpacs,amsmath,amssymb,prl,ps]{revtex4-1}
\usepackage{graphicx}
\usepackage{dcolumn}
\usepackage{bm}
\usepackage{subfigure}

\newcommand{\cdag}{c^{\dagger}}
\newcommand{\cnod}{c^{\phantom{\dagger}}}
\allowdisplaybreaks

\begin{document}

\title{Fractional quantum-Hall liquid spontaneously generated by strongly correlated  $t_{2g}$ electrons}

 \author{J\"orn  W.F. Venderbos}
 \author{Stefanos Kourtis}
 \author{Jeroen {van den Brink}}
 \author{Maria Daghofer}
 \affiliation{
  Institute for Theoretical Solid State Physics, IFW Dresden, 01171 Dresden, Germany 
 }

\date{\today}

\begin{abstract}
For topologically nontrivial and very narrow bands, Coulomb repulsion
between electrons has been predicted to give rise to a spontaneous fractional
quantum-Hall (FQH) state in absence of magnetic fields. Here we show
that strongly correlated electrons in a $t_{2g}$-orbital system on a triangular lattice
self-organize into a spin-chiral magnetic ordering pattern that 
induces precisely the required topologically nontrivial and flat
bands. This behavior is very robust and does not rely on
fine tuning. In order to go beyond mean field and to study the impact of
longer-range interactions, we map the low-energy
electronic states onto an effective one-band model. Exact
diagonalization is then used to establish signatures of a spontaneous FQH state. 
\end{abstract}

\maketitle

The Integer Quantum Hall (IQH) effect~\cite{Klitzing80} is a prime
example of an electronic state that cannot be classified within the
traditional framework of symmetry breaking, but 
is instead characterized by a topological invariant~\cite{TKNN}.
It is by now theoretically well established that an external magnetic
field is in principle not needed and that states within the same
topological class as IQH states can be realized in lattice models, if
time-reversal symmetry is broken by other mechanisms, e.g., by complex
electron hoppings~\cite{haldane88}. Related topologically nontrivial
Quantum  Spin-Hall (QSH) states even occur in systems where
time-reversal symmetry is not broken at
all~\cite{PhysRevLett.95.146802,PhysRevLett.98.106803,PhysRevB.79.195322,PhysRevB.75.121306,PhysRevLett.96.106802}, see
Refs.~\cite{RMP_Hasan_Kane,RMP_Qi_Zhang} for reviews.    
At present, many intriguing features intrinsic to topologically
non-trivial states have been observed in the absence of magnetic fields, such as the metallic Dirac
cones at the surface of a topological
insulator~\cite{topo_BiSe_2009,xia_hasan_2009}, or the QSH 
effect in quantum wells~\cite{Bernevig15122006,Konig02112007}.

Fractional Quantum Hall (FQH) states~\cite{PhysRevLett.48.1559} are
topological states that can 
be seen as composed of quasi-particles carrying an exact fraction
of the elementary electronic charge~\cite{Laughlin:1983}. 
Apart from the fundamental interest in observing  
a quasi-particle that behaves in many ways like a fraction of an
electron, some FQH states also have properties relevant to 
fault-tolerant quantum computation~\cite{RevModPhys.80.1083}.  
Very recently~\cite{tang10,sun10,neupert10}, it was suggested that lattice-FQH
states may similarly arise without a magnetic field, in fractionally
filled topologically nontrivial bands. 

In contrast to the IQH and QSH effects, which can be fully
understood in terms of non-(or weakly-)interacting electrons,
interactions are an essential requirement for FQH states, which
places demanding restrictions on candidate systems: One needs a
topologically nontrivial band that must be nearly flat  
--  similar to the highly degenerate Landau levels -- so that the
electron-electron interaction can at the same time be large compared
to the band width and small compared to the gap separating it from
other bands~\cite{tang10,sun10,neupert10}. If the
requirements can be fulfilled, however, the temperature scale of the
FQH state is set by the energy scale of the interaction. This can
allow temperatures considerably higher than the sub-Kelvin range of
the conventional FQH effect, which would  be  
extremely desirable in view of potential quantum-computing
applications. Moreover, the lattice version of FQH states~\cite{Qi:2011QAHstates}
may have unique and different properties.~\cite{2011arXiv1112.3311B}.

In most recently proposed model
Hamiltonians~\cite{tang10,sun10,neupert10,Ruby2011,Venderbos:11_flat,FQHbosons2011}, 
the topological nature of the bands was introduced by hand and model
parameters have to be carefully tuned to obtain very flat bands.  
As potential realizations, ``purpose built'' physical systems in 
oxide heterostructures~\cite{Okamoto_het2011} or optical lattices~\cite{sun10} were
suggested. On the other hand, topologically nontrivial bands can in principle 
emerge spontaneously in interacting electron systems~\cite{Mott_topo,PhysRevLett.103.046811},
e.g., for charge-ordered
systems~\cite{Rev_TI_SL:2011,PhysRevLett.104.196401} or for electrons
coupling to spins in a non-coplanar magnetic
order~\cite{Matl:1998p2624,Taguchi:2001p2556}. 
We demonstrate here that such a scenario indeed arises 
in a Hubbard model describing electrons with a $t_{2g}$ orbital degree of freedom
on a triangular lattice: a ground state with topologically nontrivial and nearly
flat bands is stabilized by onsite Coulomb interactions. Upon
doping the flat bands, longer-range Coulomb repulsion induces FQH states.

\begin{figure}
\centering
\subfigure{\includegraphics[width=0.48\columnwidth]{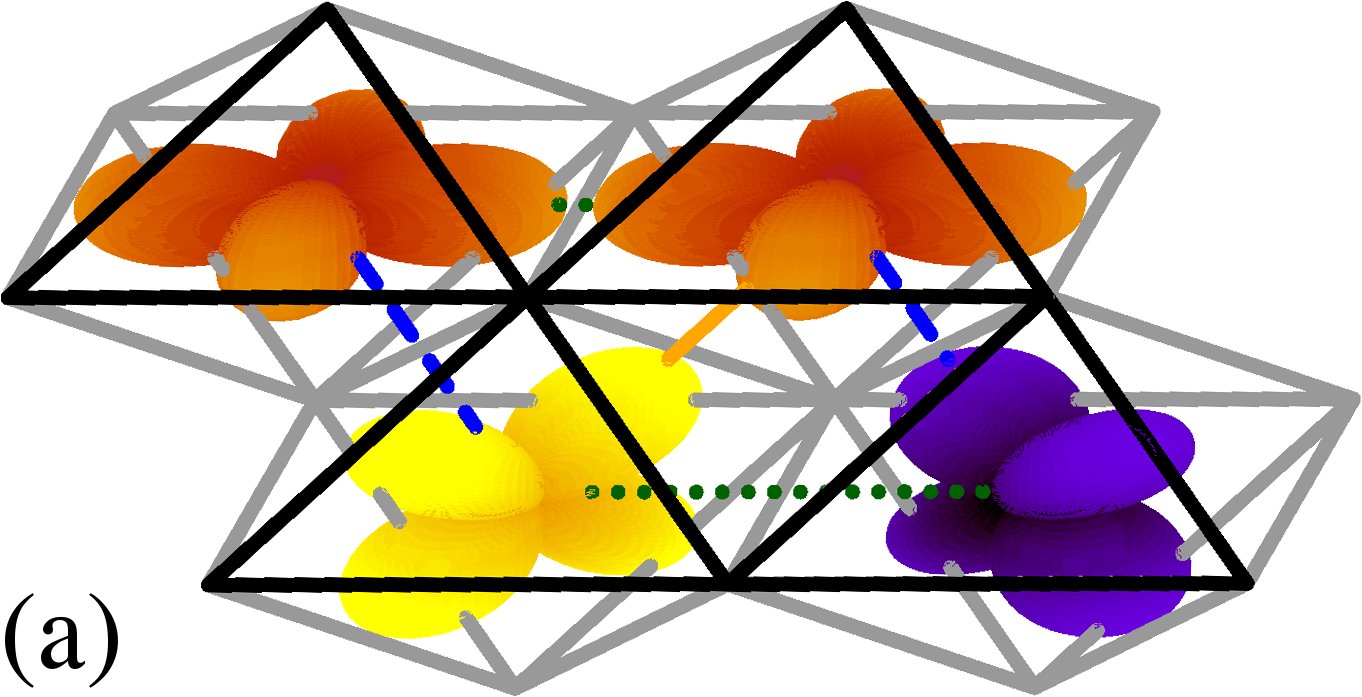}\label{fig:xy_xz_yz}}
\subfigure{\includegraphics[width=0.48\columnwidth]{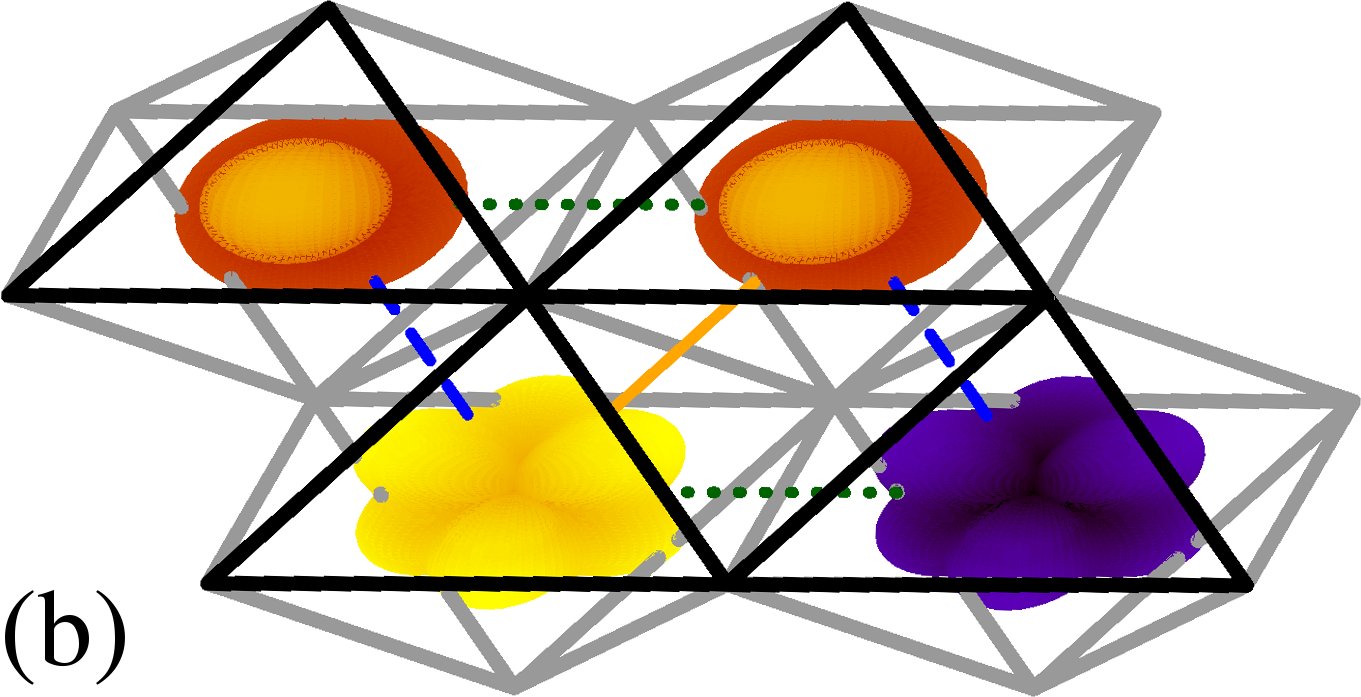}\label{fig:a1g_eg}}\\
\caption{(Color online) 
  Triangular perovskite lattice and $t_{2g}$ orbitals. Oxygen
  octahedra are indicated by lines, with black lines illustrating the
  front facets. Thick dotted (dashed, solid) lines indicate nearest-neighbor bonds
  along lattice vector ${\bf a}_1$ (${\bf a}_2$, ${\bf a}_3$). (a) Shows two $d_{xy}$
  orbitals (top) and one $d_{xz}$ and $d_{yz}$ orbital (bottom). In
  (b), the orbitals reflecting the three-fold lattice symmetry are shown: The
  two $e_g'$ orbitals (bottom), which differ by their complex phases,
  will turn out to be half filled, while the $a_{1g}$ orbital
  (pointing out of the plane, see top) forms nearly flat bands with
  non-trivial topological character that can support spontaneous FQH states.
\label{fig:t2g}}
\end{figure}

{\it $t_{2g}$ orbitals on the triangular lattice.---} 
The building blocks of our
system are oxygen octahedra with 
a transition-metal (TM) ion in the center, the most
common building block in the large and versatile material class of TM
oxides. Local cubic symmetry due to the oxygen ions splits the $d$-orbitals into 
$t_{2g}$ and $e_g$ levels, and it has been shown that orbital degrees
of freedom of either kind can substantially reduce the width of
topologically nontrivial bands~\cite{Venderbos:11_flat}. Here, we
concentrate on the $t_{2g}$ orbitals illustrated in
Fig.~\ref{fig:t2g}(a), which are further split by a crystal-field due
to the overall lattice geometry. On a triangular lattice, we find one
$a_{1g}$ and two $e_{g,\pm}^{\prime}$ 
states, see Fig.~\ref{fig:t2g}(b), with a splitting $H_{\textrm{JT}} =  \Delta_{\textrm{JT}} (n_{e_{g+}}+n_{e_{g-}}
-2n_{a_{1g}})/3$ depending on the Jahn-Teller effect and the lattice~\cite{Koshibae03}. 
Electron hopping along nearest-neighbor (NN) bonds consists of terms
$t$ via ligand oxygens and $t_{dd}$ due to direct $d$-$d$
overlap~\cite{Pen97,Koshibae03}, hopping matrices are given
in~\cite{suppl}. We set here $n<3$ and choose $t>0$~\cite{Koshibae03}
as unit of 
energy, but analogous results hold for $n>3$, $t<0$, and $t_{dd}\to
-t_{dd}$, $\Delta_{\textrm{JT}}\to -\Delta_{\textrm{JT}}$ due to
particle-hole symmetry. 

 \begin{figure}
   \centering
\begin{minipage}{0.32\columnwidth}
  \subfigure[]{\includegraphics[width=\textwidth,angle=0,trim =100 90 100 90]
    {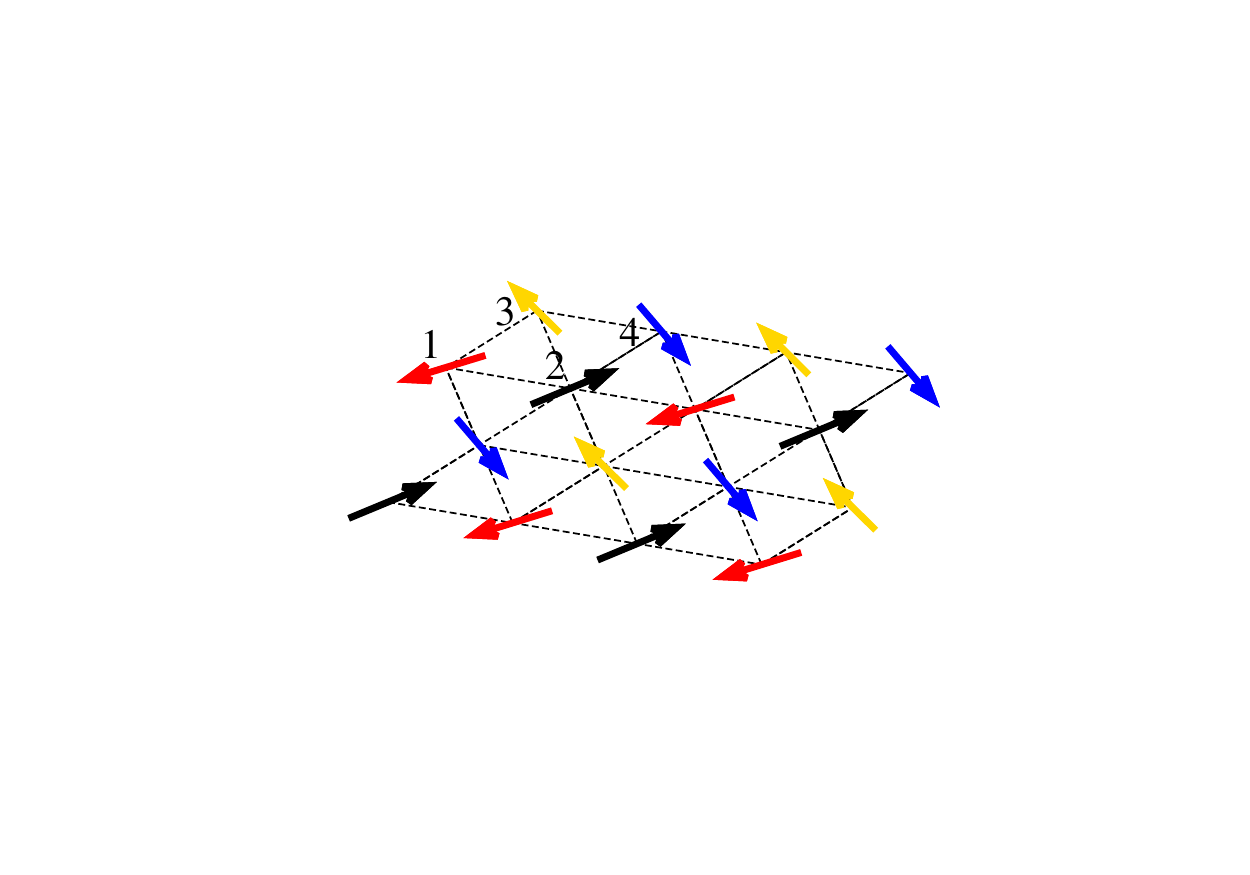}\label{fig:tri_chiral}}\\
  \subfigure[]{\includegraphics[width=0.55\columnwidth,trim =0 0 0 0 ,clip]
    {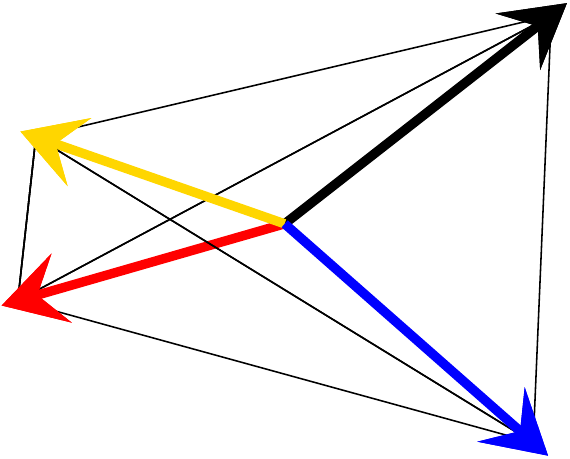}\label{fig:tetra}}
  \end{minipage}
  \begin{minipage}{0.64\columnwidth}
   \subfigure{
    \includegraphics[width=\textwidth,trim =0 0 0
    0,clip]{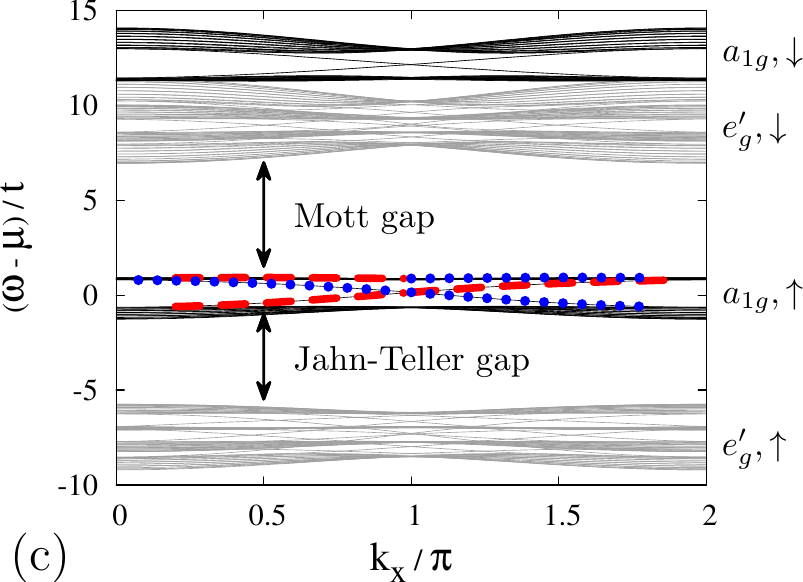} \label{fig:flatbands_hubb}}
  \end{minipage}
 \caption{(Color online) 
 Spin-chiral magnetic phase with topologically nontrivial bands
  stabilized by onsite Coulomb interactions in $t_{2g}$ electrons on a
  triangular lattice. 
  (a) Chiral magnetic order, the sites of the unit
   cell are labeled by $1$ to $4$. (b) The spins on the four sites can be
   seen as pointing to the corners of a tetrahedron, i.e., the pattern
   is non-coplanar.
   (c) One-particle energies on a cylinder 
   (periodic boundary conditions along $x$) in the mean-field~\cite{suppl} ground
   state of the $t_{2g}$ multiorbital Hubbard model, which is given by
   the pattern shown in (a). 
   States drawn in black (grey) have more (less) than $33\%$ $a_{1g}$
   character, dashed and dotted lines indicate edge states with more
   than $33\%$ of their weight on the top (bottom) row of 
   sites. The arrows $\uparrow$ ($\downarrow$) indicate states with
   electron spin mostly (anti-)parallel to the local quantization
   axis, which can be seen as the lower (upper) Hubbard band. The
   filling is $2.5$ electrons per site, slightly less than half
   filling. Parameters used were $t=1$, $t_{dd}=0$, $U/t=12$, $J/t=3$,
   $\Delta_{\textrm{JT}}/t=-6$. The figure of merit $M$, which is given by the ratio
   of the gap separating the two $a_{1g}$ subbands of the lower
   Hubbard band and the band width of the highest subband of the lower
   Hubbard band, is $M \approx 14$.
\label{fig:chiral}}
 \end{figure}

In TM oxides, Coulomb interaction is substantial compared to the
kinetic energy of $t_{2g}$ orbitals and spin-orbital physics induced
by correlations  are known to be rich in $t_{2g}$ systems on
triangular lattices~\cite{Pen97,normand:094427}. We  
take into account the onsite interaction including Coulomb repulsion
$U$ (intra-orbital) and $U'$ (interorbital) 
as well as Hund's-rule coupling $J$.  
We employ a mean-field approximation with a decoupling into
expectation values of densities   
$\langle n_{{\bf i},\alpha,\sigma}\rangle=\langle \cdag_{i,\alpha,\sigma}
\cnod_{i,\alpha,\sigma} \rangle$ for site ${\bf i}$, orbital $\alpha$, and
spin $\sigma$~\cite{Nomura:2000p647,martin08}. The spin is thus
reduced to its $z$-component $m_{{\bf  
    i},\alpha}= (n_{{\bf i},\alpha,\uparrow} - n_{{\bf
    i},\alpha,\downarrow})/2$ and non-collinear magnetic patterns are
treated by allowing for a 
site-dependent spin-quantization axis expressed by angles $\theta_i$ and
$\phi_i$. The change in quantization axis 
from site to site manifests itself in a complex Berry phase for the hopping
terms~\cite{Dagotto:Book}. Numerical optimization is used to find the
$\theta_i$ and $\phi_i$ giving the magnetic ground state, permitting arbitrary magnetic
orderings with unit cells of up to four sites, including all phases
considered in Ref.~\cite{Akagi:2010p083711}.  
For simplicity, we present here results for $J/U = 1/4$ and the
relation $U'=U-2J$ between the Kanamori parameters was used, but we
have verified that the results presented remain robust for other
choices. For details see~\cite{suppl}.

For wide parameter ranges (see below), the ground state is the non-coplanar spin-chiral
phase illustrated in Fig.~\ref{fig:chiral}(a,b). As demonstrated in
the context of the Kondo-lattice~\cite{martin08,Akagi:2010p083711} and the Hubbard~\cite{martin08,2011arXiv1103.2420L} models, 
this magnetic order leads to topologically nontrivial bands, which can
also be seen in the
one-particle bands shown in Fig.~\ref{fig:chiral}(c).  
The chemical potential lies within the $a_{1g}$ states of the lower Hubbard band, where the
electron spin is mostly parallel (labelled by $\uparrow$) to the
direction defined by the spin-chiral pattern. Dashed and dotted lines
decorate states living on the top and bottom edges of a cylinder,
they cross the chiral gap exactly once as one
left- and one right-moving edge mode, indicating the different Chern
numbers associated with the two bands directly above and below the chemical
potential. 
Such a spontaneous IQH state is already rather exotic and has recently been
shown to support fractionalized excitations bound to
vortices~\cite{2011arXiv1112.3347M}.

 \begin{figure}
\centering
   \subfigure{
\includegraphics[width=0.48\columnwidth,trim =6 15 5
     40,clip]{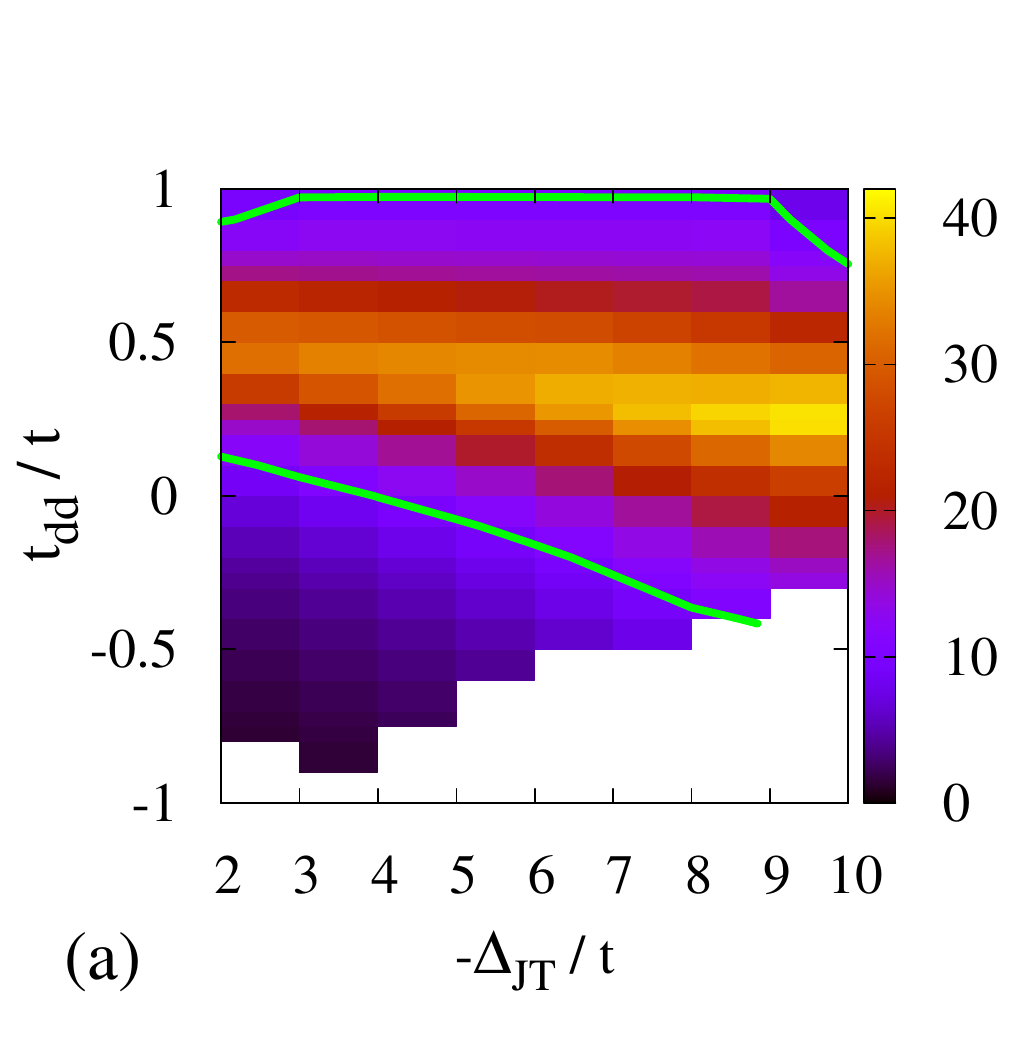} \label{fig:chiral_U12}}
   \subfigure{
\includegraphics[width=0.44\columnwidth,trim =5 15 10
     0,clip]{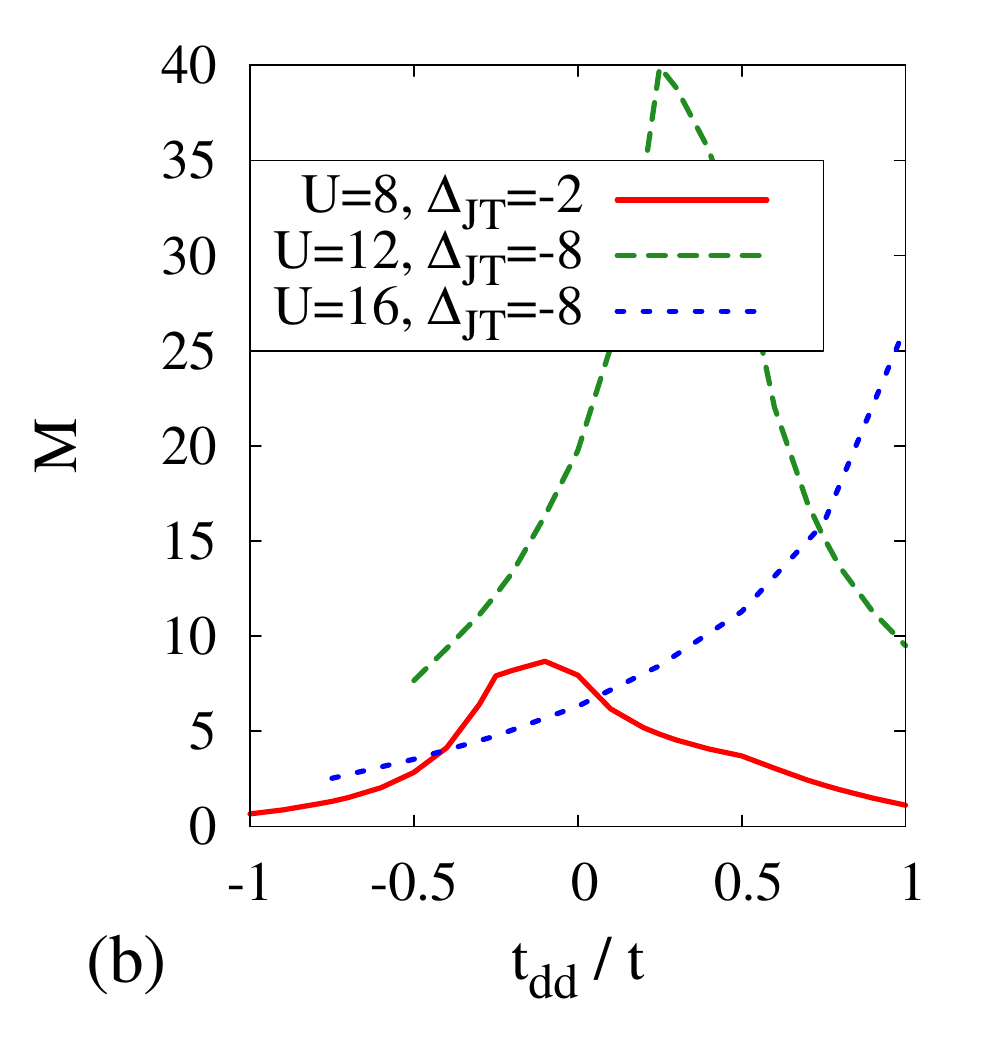} \label{fig:gap_tdd_U}}\\
 \caption{(Color online) Stability of the spin-chiral phase and flatness of the
topological bands depending on parameters of the Hamiltonian. In
(a), shaded areas in the $t_{dd}$-$\Delta_{\textrm{JT}}$ plane
indicate a spin-chiral ground  
   state Fig.~\ref{fig:chiral}(a,b) for $U/t=12$, white areas have a different
   ground state. Shading indicates the figure of merit $M$ for the
   flatness of the upper chiral subband, bright thick lines bound the
   region with $M\geq 10$. (b) shows $M$ depending on $t_{dd}$ for
selected sets of $U$ and $\Delta_{\textrm{JT}}$.  Where the
   ``Mott gap'', which separates the flat topologically non-trivial band from the upper Hubbard band, becomes
   very small, $M$ is determined by the minimal gap separating the band of interest from other
   bands. $J=U/4$ and $t=1$ were used in all cases.
\label{fig:stable}}
 \end{figure}

Figure~\ref{fig:chiral}(c)
also indicates that the upper
chiral subband has a very small width, $\sim 14$ times
smaller than the chiral gap. One can quantify the band flatness by a
figure of merit $M$ given by the ratio of the gap 
to the band width. Its dependence on various parameters of the
Hamiltonian is shown in Fig.~\ref{fig:stable}. It peaks at $M>
40$, but the more striking observation is that it is above 5 or even 10 for wide ranges
of $U$, $\Delta_{\textrm{JT}}$ and $t_{dd}$, in contrast to many other
proposals that require carefully fine-tuned
parameters~\cite{tang10,sun10,neupert10,sheng11,Ruby2011,Venderbos:11_flat,FQHbosons2011}. Nearly
flat chiral bands are thus very robust in this system and both their
topological character and their flat dispersion emerge spontaneously
with purely onsite interaction and short-range hopping, without 
spin-orbit coupling or any explicit breaking of time-reversal symmetry. 

{\it Mapping to an effective model.---}  
For large onsite interactions and large crystal field
splitting $U,J,|\Delta_{\textrm{JT}}|\gg t,t_{dd}$, the three-orbital
model with fillings between 2 and 3 electrons per site 
can be mapped onto the one-band Kondo-lattice model (KLM). Low-energy
configurations minimize onsite interactions and thus 
contain two or three electrons per site, with parallel spins due to Hund's
rule. In order to additionally minimize the crystal-field energy, the $e_g'$
levels will always be half filled and form an effective spin, 
while any holes will be found in the $a_{1g}$ sector.  The 
electrons in the partially filled $a_{1g}$ states can delocalize with
an isotropic hopping $t_{a1g}=(2t+t_{dd})/3$, however, their spin must
remain parallel to the local $e_g'$ spin. In the low-energy limit, each
site can thus be described as a spin coupled to a charge degree of
freedom and we arrive at the situation
described by the KLM in the limit of strong Hund's rule coupling. Our
numeric mean-field results corroborate this picture, see
Fig.~\ref{fig:chiral}(c), where the $e_g'$ levels are found far below
the chemical potential. The KLM supports spin-chiral phases on many
frustrated lattices like the 
triangular~\cite{martin08,Akagi:2010p083711,Kato_FKLM_tri_2010,Kumar:2010p216405},
pyrochlore~\cite{Chern10}, and face-centered cubic~\cite{Shindou01} lattices.

In addition to processes within the low-energy Hilbert space, virtual
excitations involving high-energy states can be taken into account in
second-order perturbation theory. This leads to (i) effective
longer-range hopping of the $a_{1g}$ 
electrons 
and (ii) an effective
antiferromagnetic superexchange between the $e_g'$ spins. The latter 
stabilizes the spin-chiral pattern~\cite{Kumar:2010p216405} and is due to excitations into the
upper Hubbard/Kondo band. When it is suppressed for $U\gtrsim 24 |t|$,
the ground state consequently becomes FM, as in 
the KLM with a large Kondo
gap~\cite{Kato_FKLM_tri_2010,Akagi:2010p083711}. 
Nevertheless, the
exotic spin-chiral state is remarkably stable in the present $t_{2g}$
system considering its sensitivity to Hund's coupling in the
KLM~\cite{Akagi:2010p083711}. 

The effective longer-range hopping of $a_{1g}$ electrons involves
processes via excitations into the upper Kondo/Hubbard band ($\propto
1/J$ and $\propto 1/U$) as well as virtual excitations of $e_g'$ electrons into $a_{1g}$
states ($\propto 1/\Delta_{\textrm{JT}}$), for details
see~\onlinecite{suppl}. Second-neighbor hopping $\propto 1/J$ does not
significantly modify the low-energy bands and drops out completely in the
limit of a large Mott/Hubbard gap, but third-neighbor hopping $t_3$ is
crucial in cancelling the dispersion coming from NN hopping $t_1$ for
one of the bands~\cite{suppl}. The simplest description of the
effective low-energy bands around the Fermi level is thus 
\begin{align}\label{eq:eff}
H_{\textrm{eff}}({\bf k}) &= 2t_{1} \sum\nolimits_j \sigma^j \cos {\bf k} {\bf
    a}_j + 2t_3  \sum\nolimits_j \sigma^0 \cos 2{\bf k} {\bf a}_j\;,
\end{align}
where ${\bf a}_j$ ($j=1, 2, 3$) denote the unit vectors on the triangular 
lattice. 
%
Pauli matrices $\sigma^j$ and unit matrix 
$\sigma^0$ refer to the two sites of the electronic unit
cell in the chiral phase~\cite{martin08}. Formally, this describes
electrons moving in a constant (and very strong) magnetic field with a
flux of $\pi/2$ threading each triangle of the lattice~\cite{martin08}.

 \begin{figure}
   \centering
   \includegraphics[width=\columnwidth]{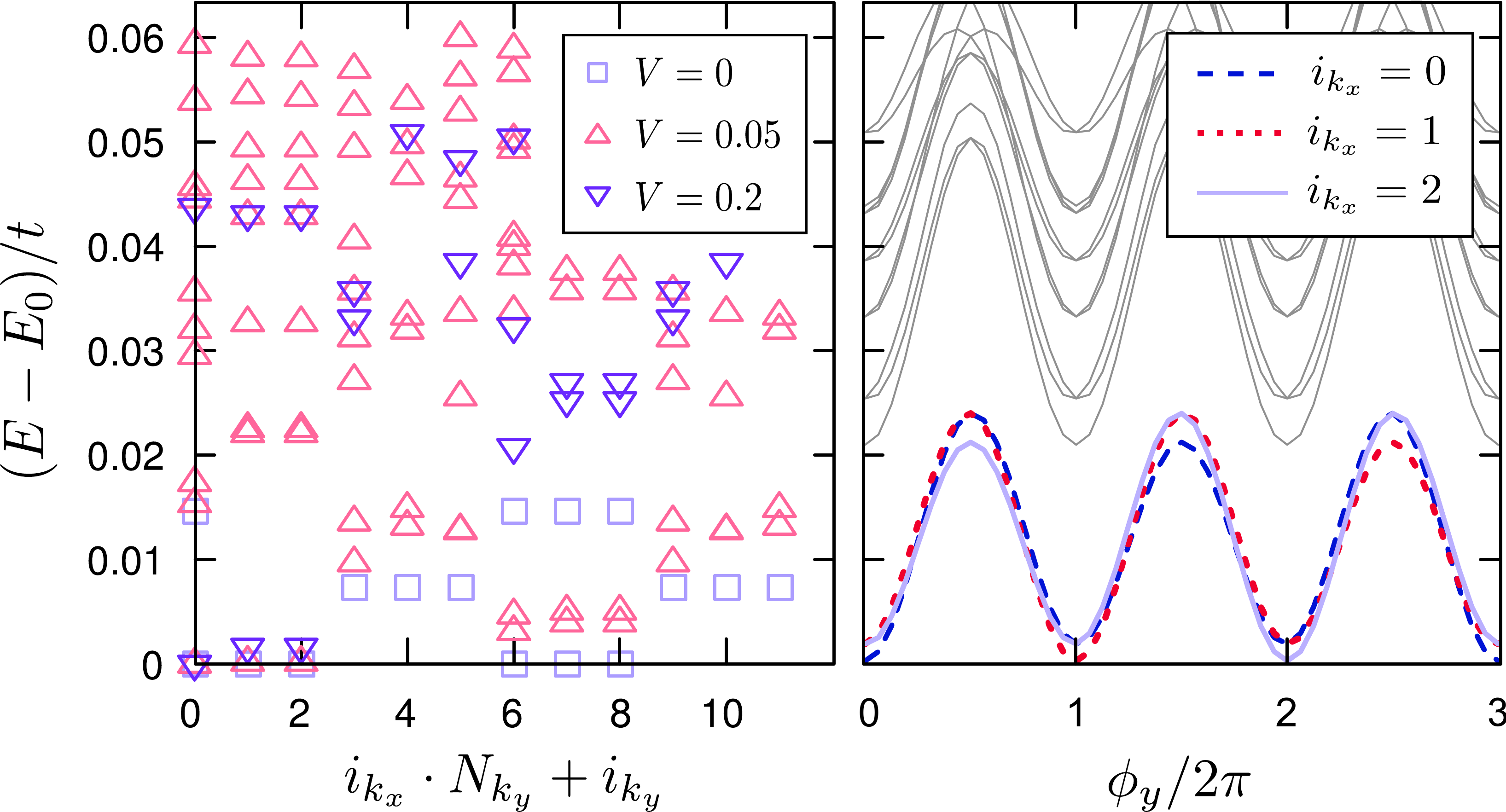}
 \caption{(Color online)  FQH state induced by NN Coulomb repulsion $V$ in the effective one-band
   model Eq.~\ref{eq:eff}. (a) Energy depending on total
   momentum ${\bf k}$ for several values of $V/t$.
   (b) Energy for $V/t=0.2$ depending on a flux $\phi_y$ added whenever an
   electron goes once around the whole lattice in $y$ direction. Each addition of
   $\phi=2\pi$ leads to an equivalent state, $6\pi$ to the same
   state. The Chern numbers associated with the three low-energy
   states are almost exactly $2/3$  for $V/t=0.2$. Lattice size is
   $4\times 6$ sites (12 two-site unit cells), 
   parameters in Eq.~\ref{eq:eff} are $t_{1}= 0.27t$ and $t_3 =
   -0.06t$, giving bands with $M\approx 13$ and a gap of
   $0.89t$. The filling of the flat band is $2/3$. \label{fig:FQHE}} 
 \end{figure}

{\it FQH groundstates of an effective spinless one-band model.---} 
We now address the impact of NN Coulomb interaction $V\sum_{\langle
  i,j \rangle}n_in_j$ on the fractionally filled flat band. 
The spin-chiral state can only be expected to remain stable for
densities close to 2.5 electrons per site, i.e., low
doping factions $\nu$ of the flat band~\cite{Akagi:2010p083711}. 
FQH states corresponding to such low fillings are generally separated
from the rest of the spectrum by only a small gap, making their
analysis on finite-size clusters difficult~\cite{Regnault2011}. Here, we use  Lanczos exact
diagonalization~\cite{neupert10,sheng11,Regnault2011,suppl} to study
a number of simple filling fractions ($1/3$, $2/3$,
$1/5$, and $2/5$) available on accessible lattices and consistently
find $V$ to induce signatures of a FQH 
state. It is thus plausible that the FQH behavior discussed next
persists to fractional fillings in a low doping range of the
spin-chiral state. 

As an example, we present here the case of 16
electrons on a $4\times 6$-site cluster of the model
Eq.~(\ref{eq:eff}), a filling that would correspond to $2.\dot{6}$ in
the original three-orbital model. After a particle-hole
transformation, it corresponds to $2/3$ filling of the nearly flat
band.  
Figure~\ref{fig:FQHE}(a) shows that with increasing $V$, three
low-energy states split off from the rest of the spectrum. Inserting a
magnetic flux $\phi_y=2\pi$ interchanges the three 
states, $\phi_y=6\pi$ recovers the original situation, see
Fig.~\ref{fig:FQHE}(b), as reported for  
other systems~\cite{neupert10,sheng11,Venderbos:11_flat}. The Chern
number $C$ is evaluated by integrating the
flux-dependent Berry curvature $\Omega^n(\phi_x,\phi_y)$ (obtained by the
Kubo formula~\cite{Niu:1985sigmaxy,Xiao:2010Berry,suppl}) over the square $0\leq
\phi_x, \phi_y< 6\pi $. For $V=0.2|t|$, the three low-energy states
have Chern numbers within $1\%$ of the expected $C=2/3$, a deviation
well within the limits of reported finite-size effects~\cite{neupert10}. 

{\it Conclusions.---} The possibility of a spontaneous FQH effect without a magnetic field is
currently hotly discussed, and various 
models have been
suggested~\cite{tang10,sun10,neupert10,sheng11,Ruby2011,Venderbos:11_flat,FQHbosons2011,Okamoto_het2011}.  
However, an experimental realization appears challenging, as the
necessary topological character and the flatness of the bands need to
be carefully engineered in previous proposals. 
We have shown here that  bands with the desired
properties emerge spontaneously for wide parameter ranges in strongly
correlated $t_{2g}$ orbitals on a triangular lattice, and that
these bands support FQH ground states. Both $t_{2g}$ systems and triangular lattices occur in
various TM oxides, and signatures of the
unconventional \emph{integer} QH state have been reported for a triangular-lattice palladium-chromium
oxide~\cite{AHE_PdCrO2_2010}. This harbors the prospect that a suitable
material can be synthesized in this highly versatile material class. As 
such a material is by default strongly correlated, 
one also naturally expects an inter-site Coulomb repulsion that is
strong enough to stabilize spontaneous FQH states in the absence of a
magnetic field. 

\acknowledgments
This research was supported by DFG (Emmy-Noether
program; SK and MD) and the Interphase Program of the Dutch
Science Foundation NWO/FOM (JV and JvdB). 

\appendix
\vspace*{2em}
{\it The appendix contains Supplemental Material:}

\section{One-particle terms of the multi-orbital $t_{2g}$ Hamiltonian}

The multi-orbital kinetic energy is 
\begin{align}\label{eq:ekin_gen}
H_{\rm kin} = 
\sum_{\langle i,j\rangle,\alpha,\beta,\sigma} 
t_{i,j}^{\alpha,\beta}
\cdag_{i,\alpha,\sigma\phantom{\beta}\hspace{-0.5em}}\cnod_{j,\beta,\sigma} + \textrm{H.c}\;,
\end{align}
where $\cdag_{i,\alpha,\sigma}$ ($\cnod_{i,\alpha,\sigma}$) creates
(annihilates) an electron on site $i$, in orbital $\alpha$ and with
spin $\sigma$. $\langle i,j\rangle$ denotes nearest-neighbor (NN) bonds, $\alpha$
and $\beta$ denote the orbital. Using  as basis states the $xy$, $xz$, and $yz$
orbitals shown in Fig.~1(a) of the main text, the orbital-
and direction-dependent hopping parameters $t_{i,j}^{\alpha,\beta}$ are
given by the matrices
\begin{gather*} 
 \hat{T}_{1} =  \begin{pmatrix}
t_{dd} & 0 & 0\\
0  & 0 & t \\
0  & t & 0 \\
\end{pmatrix}, \ 
\hat{T}_{2} = \begin{pmatrix}
0  & 0 & t\\
0  & t_{dd}& 0 \\
t  & 0 & 0 \\
\end{pmatrix}, \ 
\hat{T}_{3} =  \begin{pmatrix}
0  & t & 0\\
t  & 0 & 0 \\
0  & 0 & t_{dd}\\
\end{pmatrix}
\end{gather*}
for NN bonds along the three directions ${\bf a}_1$, ${\bf a}_2$,
${\bf a}_3$ as illustrated in Fig.~1 of the main text. The
transformation into the $\{a_{1g}, e_{g,+}^{\prime},e_{g,-}^{\prime}\}$ can be found, e.g., in
Ref.~\cite{Koshibae03}. 

\section{Mean-field approximation}

Onsite interaction is described by Kanamori parameters $U$ ($U'$) for
Coulomb repulsion between electrons in the same (different) orbitals
as well as ferromagnetic Hund's-rule coupling between electrons in
different orbitals. The relation $U'=U-2J$ is used here,
``pair-hopping'' $J'$ is left out, because it drops out of the mean
field decoupling 
\begin{align}  \label{eq:Hcoul_mf}
  H_{\rm int}& \approx 
  U\sum_{{\bf i},\alpha} \left(
\langle n_{{\bf i},\alpha,\uparrow}\rangle n_{{\bf i},\alpha,\downarrow}
+ n_{{\bf i},\alpha,\uparrow} \langle  n_{{\bf i},\alpha,\downarrow}\rangle\right)\nonumber\\
&\quad +(U'-J/2)\sum_{{\bf i},\alpha < \beta}\left(
\langle n_{{\bf i},\alpha}\rangle n_{{\bf i},\beta}
 +n_{{\bf i},\alpha} \langle n_{{\bf i},\beta} \rangle \right)\nonumber\\
&\quad -2J\sum_{{\bf i},\alpha < \beta}\left(
\langle m_{{\bf i},\alpha} \rangle m_{{\bf i},\beta}
+m_{{\bf i},\alpha} \langle m_{{\bf
    i},\beta} \rangle  \right)\nonumber\\
&\quad - U\sum_{{\bf i},\alpha} 
\langle n_{{\bf i},\alpha,\uparrow}\rangle \langle n_{{\bf i},\alpha,\downarrow}\rangle
-(U'-J/2)\sum_{{\bf i},\alpha < \beta}
\langle n_{{\bf i},\alpha}\rangle \langle n_{{\bf i},\beta}\rangle\nonumber\\
&\quad +2J \sum_{{\bf i},\alpha < \beta}
\langle m_{{\bf i},\alpha} \rangle \langle m_{{\bf
    i},\beta}\rangle, 
\end{align}
where ${\bf i}$ labels the site, $\alpha$ and $\beta$
orbitals. $n_{{\bf i},\alpha,\sigma}=\cdag_{i,\alpha,\sigma}
\cnod_{i,\alpha,\sigma}$ is the density operator. We keep here only
expectation values for diagonal operators, i.e., only 
$\langle n_{{\bf i},\alpha,\sigma}\rangle=\langle \cdag_{i,\alpha,\sigma}
\cnod_{i,\alpha,\sigma} \rangle$~\cite{Nomura:2000p647,martin08},
which reduces the spin to its $z$-component $m_{{\bf
    i},\alpha}= (n_{{\bf i},\alpha,\uparrow} - n_{{\bf
    i},\alpha,\downarrow})/2$. In
order to treat non-collinear spin patterns, one has to allow for a
site-dependent spin-quantization axis given by angles $\theta_i$ and
$\phi_i$. The change in quantization axis 
from site to site manifests itself in a complex phase for the hopping
terms,\cite{Dagotto:Book} 
which is between sites $i$ and $j$
\begin{align}\label{eq:berry}
\Omega^{\sigma,\sigma}_{ij}&=\cos \frac{\theta_{i}}{2} \cos \frac{\theta_{j}}{2}
+\sin \frac{\theta_{i}}{2} \sin \frac{\theta_{j}}{2}\textrm{e}^{-i\sigma(\phi_{i}-\phi_{j})}\nonumber\\
\Omega^{\uparrow,\downarrow}_{ij}&=\cos \frac{\theta_{i}}{2}\sin \frac{\theta_{j}}{2}\textrm{e}^{-i\phi_{j}} 
                              -\cos \frac{\theta_{j}}{2}\sin \frac{\theta_{i}}{2}\textrm{e}^{-i\phi_{i}}
\end{align}
where $\Omega^{\uparrow,\uparrow}_{ij}$
($\Omega^{\downarrow,\downarrow}_{ij}$) modulates the hopping of an electron with spin
parallel (antiparallel) to the chosen spin-quantization axis. In the
site-dependent quantization, spin is not conserved and there are
spin-mixing hoppings with $\Omega^{\downarrow,\uparrow}_{ij}$ given by the
complex conjugate of $\Omega^{\uparrow,\downarrow}_{ji}$. 

We use numerical
optimization routines to find the spin pattern with the lowest energy
among all orderings with unit cells of up to four sites, including all patterns
considered in Ref.~\cite{Akagi:2010p083711} of the main text.
In each step, the mean-field energy
is calculated self-consistently for a lattice of $16\times 16$ (four-site unit cell) or
$24\times 16$ (three-site unit cell). (For selected points in parameter
space, we also used larger lattices and did not find a significant
difference.) In order to minimize the impact of our 
approximations on the symmetries of the orbital degrees of freedom, we
perform the mean-field decoupling in the  $\{a_{1g}, e_{g,+}^{\prime},
e_{g,-}^{\prime}\}$ basis, where the symmetry between the half-filled
$e_{g,+}^{\prime}$ and the quarter-filled $a_{1g}$ orbitals (for the
fillings discussed here) is already broken by the crystal
field. We verified that decoupling directly in the $\{xy, xz,yz\}$
basis, where all three orbitals have the same electronic density, 
leads to qualitatively identical and quantitatively very similar results.

\section{Effective one-band model and exact diagonalization}

The mapping to the effective one-band model is most easily carried out in the
Kondo-lattice picture, where the localized spins are assumed to
consist of the $e_g'$ electrons. Without a magnetic order, the $a_{1g}$ orbital has an isotropic
hopping $t_{a1g}=(2t+t_{dd})/3$, i.e., the same along all three directions on the triangular
lattice, but in the spin-chiral phase, this hopping is modulated by a
direction-dependent Berry phase Eq.~(\ref{eq:berry}). The electronic
unit cell of the spin-chiral pattern has two sites~\cite{martin08},
and the Berry phases can then be expressed in terms of Pauli matrices 
as given in the main text. The absolute value of the NN hopping is
renormalized to $t_{1}=(2t+t_{dd})/3\sqrt{3}$. 

\begin{figure}
\includegraphics[width=\columnwidth]{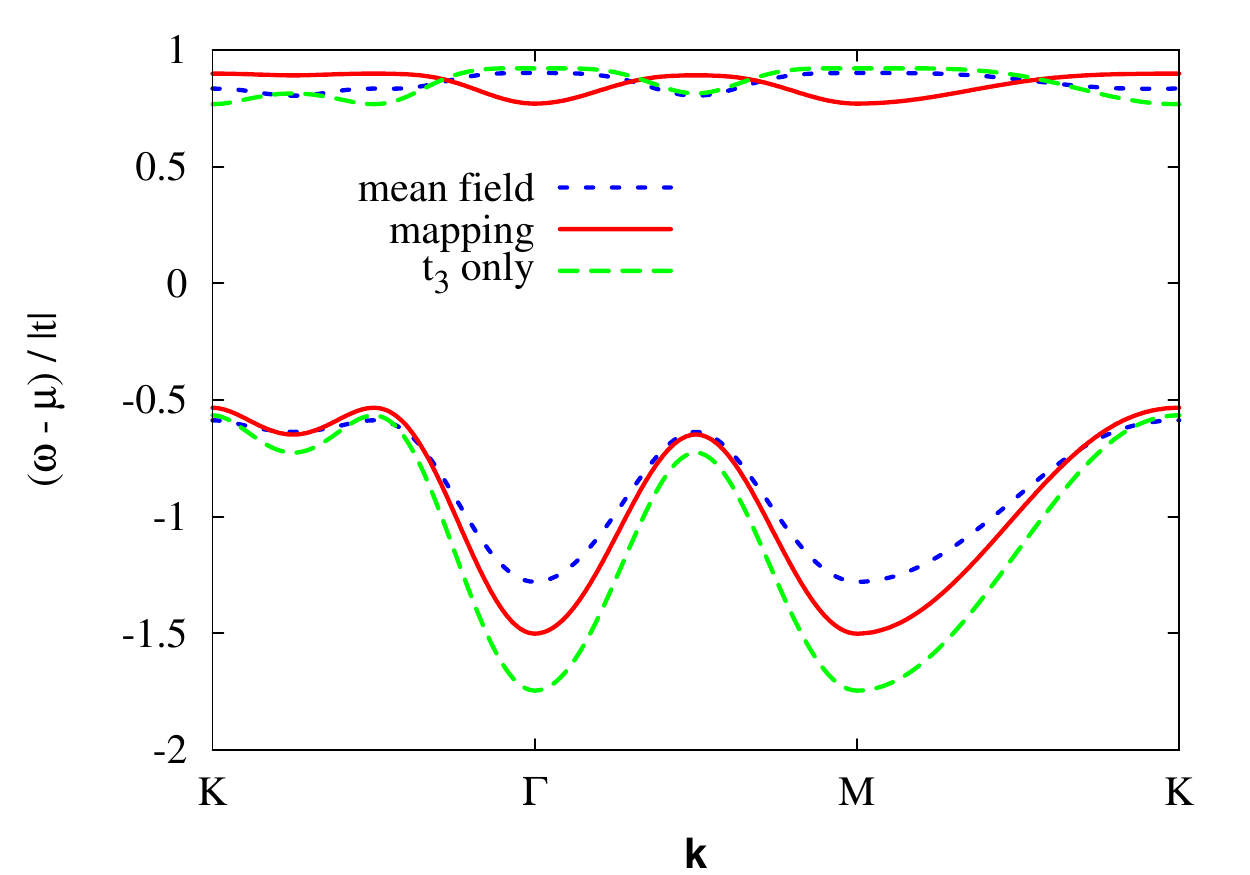}
\caption{(Color online) Comparison of the low-energy bands obtained
  in mean field theory for the spin-chiral state to those obtained in
  the strong-coupling Kondo-lattice model,
  including effective hoppings in second-order perturbation
  theory. Parameters in mean field are $U/t=12$, $J/t=3$,
  $\Delta_{\textrm JT}/t=-6$, and $t_{dd}=0$, as in Fig.~2 of the
  main text. The different dispersions in the upper band are due to the fact
  that the perturbation theory assumes that the second-order--hopping
  path is always available. As the $a_{1g}$ level is partly filled,
  some of these paths are blocked by the Pauli principle, this
  effect is included correctly in mean field. Taking this into
  account phenomenologically by reducing $t_3$ by $20\%$ improves the
  fit. This was done for the curve referred to as ``$t_3$ only'', where only
  nearest and third-neighbor hopping were included, while
  second-neighbor processes were dropped.\label{fig:mapping}}  
\end{figure}

Corrections to this simplest approximation
can be obtained by second-order perturbation theory, which yields   
longer-range hopping processes mediated by virtual excitations. We are
first going to discuss processes within the lower Kondo/Hubbard band,
where an electron from the $e_g'$ levels, which are half filled in the
low-energy Hilbert space, is excited into an empty $a_{1g}$ state in
the virtual intermediate state, involving an excitation energy
$\Delta_{\textrm JT}$. In a second step, an electron from a different 
\emph{occupied} $a_{1g}$ state can take the empty place in the $e_g'$
orbital, which corresponds to an effective hopping. There are two
possible hopping paths connecting pairs of either nearest or 
next-nearest neighbor (NNN) sites and it turns out that the corresponding effective
hoppings drop out  in the spin-chiral phase because the Berry phases
for the two paths interfere destructively. Third-neighbor sites,
on the other hand, are only connected by a single 
path and the combined Berry phase
$\Omega^{\uparrow,\uparrow}_{ij}\Omega^{\uparrow,\uparrow}_{jk}$
only renormalizes the effective hopping by a
factor of 3, because the spins at sites $i$ and $k$ are parallel in
the spin-chiral phase. One thus obtains an effective hopping
$t_3=-2(t-t_{dd})^2/(27\Delta_{\textrm{JT}})$, which is the only second-order
correction for the limit of infinite Hund's rule. In this limit, $t_3$
flattens the \emph{lower} of the subbands for the fillings discussed here~\cite{Venderbos:11_flat}. 

In the more realistic case of strong but finite Hund's rule coupling,
there are additional processes where the virtual excitation involves
an electron in the upper Hubbard/Kondo band. The corresponding
excitation energies then contain Hund's coupling $J$ and the effective
hoppings from site $i$ to $k$ involve the Berry phases
$\Omega^{\uparrow,\downarrow}_{ij}\Omega^{\downarrow,\uparrow}_{jk}$,
see Eq.~(\ref{eq:berry}). For the parameters of Fig.~2 of the main
text, the chiral bands around the Fermi level are compared to this
second-order treatment in Fig.~\ref{fig:mapping}.
NNN hopping does here not drop out, and NN
hopping is also slightly renormalized, however, both these processes
have only a small impact on the low-energy bands. Again, we find
third-neighbor hopping $t_3$ to flatten the dispersion of one of the
bands. Since excitations into the upper Hubbard/Kondo band involve
\emph{electrons} rather than holes as before, the sign of the
effective $t_3$ is reversed, and it is the upper chiral subband that
is flattened. 

In order to investigate the FQH groundstate, we used Lanczos  exact
diagonalization to study the Hamiltonian given by the kinetic
energy of the effective one-band model, Eq. (1) of the main text, and
NN Coulomb repulsion $V\sum_{\langle i,j \rangle}n_in_j$. Hopping
parameters $t_1 = 0.27$ and $t_3=-0.058$ were used, giving a
dispersion similar to Fig.~\ref{fig:mapping}. NN bonds $\langle
i,j \rangle$ are defined on the original triangular lattice and $V$
acts both between the two sites within one unit cell and between NN sites
belonging to different unit cells. As mentioned in the main text, the
spin-chiral state can actually only be expected to be stable for low
doping of the flat band~\cite{Akagi:2010p083711}, which is close to
half filling for the effective one-band model. FQH states
corresponding to such small fillings $\nu$ tend to have smaller gaps
than those for large $\nu$, and the low-energy manifold giving the
quasi-degenerate FQH states contains more states. On the small clusters that we can study
with exact diagonalization, eigenenergies always have spacings between
them, as an illustration see the $V=0$ energies in Fig.~4(a) of the
main text, which would form a continuous band in the thermodynamic
limit. It is thus far harder to reliably resolve a small gap than a
larger one, and it is moreover highly desirable that we can study the
system on at least two lattice sizes in order to see a gap. This
severely restricts our access to very low dopings. We thus study
several filling fractions corresponding to ``simple'' FQH states. In
all cases where we find a low-energy manifold to separate from the
rest of the spectrum, the states of this low-energy manifold shows
signatures of FQH behavior, which is thus a very robust feature of the
doped flat band.

Inserting a flux $(\phi_x,\phi_y)$ means that electrons gain a phase
$\textrm{e}^{i\phi_x}$ ($\textrm{e}^{i\phi_y}$) for going once around
the whole lattice in $x$-($y$-) direction. This is implemented by
changing the hopping $t_{\bf{i},\bf{j}}$ from site ${\bf i}= i_x{\bf
  a_1}+i_y{\bf a_2}$ to site ${\bf j}= j_x{\bf a_1}+j_y{\bf a_2}$  
to 
\begin{equation}
t_{\bf{i},\bf{j}} \to t_{\bf{i},\bf{j}}\textrm{e}^{i\left(\phi_x
  \frac{j_x-i_x}{L_x}+ \phi_y \frac{j_y-i_y}{L_y}\right)}, 
\end{equation}
leading to a flux-dependent
Hamiltonian $H_{\textrm{eff}}(\phi_x,\phi_y)$. In the case of
$\nu=1/3$, we find three low-energy states
separated from the remaining spectrum by a gap as in the $\nu=2/3$
case discussed in the main text, both on a  $6\times 6$ and a
$4\times 6$ system. For  $6\times 6$ sites, however, all three
low-energy states have total momentum $(0,0)$ for $(\phi_x,\phi_y)=(0,0)$. Due to finite-size
effects, the states do then not cross upon flux
insertion~\cite{Regnault2011}, but avoid crossings. For the smaller
$4\times 6$ system, the three 
low-energy states have different total momenta, and this good quantum
number allows us to clearly resolve their crossing when we insert a
flux $\phi_y$, even on a finite system. 

The Chern numbers were evaluated by integrating the Berry curvature
$\Omega^n(\phi_x,\phi_y)$ over the square $0\leq
\phi_x, \phi_y< 6\pi $.
$\Omega^n(\phi_x,\phi_y)$ was obtained by the
Kubo formula~\cite{Niu:1985sigmaxy,Xiao:2010Berry} 
\begin{widetext}
\begin{equation}\label{eq:chern}
 \Omega^n(\phi_x,\phi_y) = i L_x L_y \sum_{n'\not=n} \frac{ 
   \langle n | \frac{\partial { H}_{\textrm{eff}}(\phi_x,\phi_y)}{\partial \phi_x} |
     n' \rangle 
   \langle n' | \frac{ \partial { H}_{\textrm{eff}}(\phi_x,\phi_y)}{ \partial
     \phi_y} | n\rangle -  
    \langle n
     |  \frac{ \partial { H}_{\textrm{eff}}(\phi_x,\phi_y)}{ \partial \phi_y} |
     n'\rangle 
     \langle n' | \frac{ \partial { H}_{\textrm{eff}}(\phi_x,\phi_y) }{ \partial
     \phi_x} | n \rangle}
{(\epsilon_n -
   \epsilon_{n'})^2}, 
\end{equation}
\end{widetext}
where $n'$ and $n$ label eigenstates with energies $\epsilon_{n'/n}$ and $\partial {
  H}_{\textrm{eff}}(\phi_x,\phi_y)/\partial \phi_{x/y}$
are current operators.


%

\end{document}